\documentclass[aps,preprint,amssymb,superscriptaddress]{revtex4-1}
\usepackage{floatrow}
\newfloatcommand{capbtabbox}{table}[\FBwidth]
\usepackage{graphicx}% Include figure files
\usepackage{dcolumn}% Align table columns on decimal point
\usepackage{bm}% bold math
\usepackage{natbib}
\usepackage{CJK}
\usepackage[utf8]{inputenc}
\usepackage[T1]{fontenc}
\usepackage{mathptmx}
\usepackage{gensymb}
\usepackage{graphicx}% Include figure files
\usepackage{epstopdf}
\usepackage{dcolumn}% Align table columns on decimal point
\usepackage{bm}% bold math
\usepackage{amsmath}
\usepackage{mathtools}
\usepackage{relsize}
\usepackage{multirow}
\usepackage[colorlinks]{hyperref}
\begin{document}

%\preprint{AIP/123-QED}

%\begin{CJK*}{GB}{}
%\preprint{APS/123-QED}
\title{Computing Using Shallow N$V$-Center Charges in Diamond}% Force line breaks with \\

\author{Rodrick Kuate Defo}
\email{rkuatede@syr.edu}
\affiliation{Department of Electrical Engineering and Computer Science, Syracuse University, Syracuse, NY 13210}
\affiliation{Department of Electrical and Computer Engineering, Princeton University, Princeton, NJ 08540}
\author{Steven L. Richardson} 
\affiliation{John A. Paulson School of Engineering and Applied Sciences, Harvard University, Cambridge, MA 02138}
\affiliation{Department of Electrical and Computer Engineering, Howard University, Washington, DC 20059}
\date{\today}% It is always \today, today,

\begin{abstract}
The static electric dipole-dipole coupling between donor-acceptor pairs (DAPs) in wide-bandgap semiconductors has recently emerged as a means of realizing a quantum science platform through optically controllable, long-range interactions between defects in the solid state. In this work, we generalize DAPs to consider arbitrary dopant populations and demonstrate that the charge of the N$V$ center in diamond is well suited for quantum science. Explicitly, we leverage experimental results [see Z. Yuan \textit{et al}., PRR 2, 033263 (2020)] to show that shallow N$V$ centers can be efficiently initialized to a given relative population of the negative and neutral charge states and that modulating the surface termination would allow for control of the timescale over which the initialization and subsequent computations would occur. Furthermore, we argue that the observation of electroluminescence from the neutral charge state of the N$V$ center [see N. Mizuochi \textit{et al}., Nat. Photon. 6, 299 (2012)], but not from the negative charge state, implies the ability to interface with the N$V$ center's charge in a manner analogous to the spin interface enabled by the spin-state dependent fluorescence of the N$V$ center.

\end{abstract}

\maketitle
%\preprint{APS/123-QED}

\section{INTRODUCTION}
Nonradiative charge-carrier recombination at defect sites, known as Shockley-Read-Hall recombination~\cite{abakumov1991nonradiative,Shockley1952statistics,Hall1952electron}, decreases the efficiency of optoelectronic devices~\cite{Alkauskas2014first}, in particular by reducing the luminescence, carrier lifetimes, and number of photogenerated carriers~\cite{Shi2015comparative,Stoneham1981non}. To address the issue of nonradiative charge carrier recombination, Henry and Lang developed a theory of nonradiative capture by multiphonon emission in their seminal paper~\cite{Henry1977nonradiative}, building on an earlier theory by Marcus dealing with charge transfer based on the fluctuations of various nuclear coordinates~\cite{Marcus1993electron,Shi2015comparative}. Later works have applied advances in computational capabilities to transform the theory of Henry and Lang into an \textit{ab initio} formalism by explicitly calculating the electron-phonon coupling matrix elements involved in the Fermi's golden rule formulation of capture rates~\cite{Alkauskas2014first,Shi2015comparative,Turiansky2021nonrad,Barmparis2015theory}. Some of these works have used the approximation of employing only one special phonon mode to replace the sum over all vibrational degrees of freedom in the rate calculations ~\cite{Alkauskas2014first,Turiansky2021nonrad} while others have argued that the full summation over all vibrational degrees of freedom cannot be avoided~\cite{Shi2015comparative,Barmparis2015theory}. We argue in this paper that the full summation over all vibrational degrees of freedom would not be necessary in the isotropic limit, or equivalently in the limit where averaging over sufficiently many randomly or uniformly distributed dopants can be performed. In that limit, a single phonon state (potentially constructed from a superposition of phonon modes) is required; however, there is no need to explicitly calculate the phonon modes from which the phonon state is constructed as we will show below. Leveraging this formalism, we propose a mechanism for the initialization of the N$V$ center's charge state that would be suitable for computing with arbitrary dopant populations. We also argue that the timescale over which the initialization of the N$V$ center's charge state and subsequent computations would occur could be controlled by the surface termination of the host, by adsorbing or desorbing atoms on the crystal, or through scanning tunneling microscopy (STM). Finally, we argue that charge-state dependent electroluminescence of the N$V$ center in diamond~\cite{Mizuochi2012} would serve an analogous role to the spin-state dependent fluorescence of the N$V$ center in allowing for an optical interface. We note that our formalism applies equally well to group-IV impurity-vacancy color centers in diamond such as the Si$V$~\cite{Zhang2023Neutral,neu2011single,Kuate2021calculating,haussler2017photoluminescence,ekimov2019effect},  Ge$V$~\cite{Ekimov2015germanium,Kuate2021calculating,Kuate2019how,Iwasaki2015germanium,Ralchenko2015observation,Ekimov2017anharmonicity,Palyanov2016high,Palyanov2015germanium,Bhaskar2017quantum,bray2018single,Harris2023hyperfine}, Sn$V$~\cite{Iwasaki2017tin,Kuate2021calculating,Ekimov2018tin,tchernij2017single,palyanov2019high, Alkahtani2018tin, rugar2019char, wahl2020direct,fukuta2021sn,Gorlitz2020spectroscopic,Trusheim2020transform,Rugar2021quantum,Harris2023hyperfine,Parker2024a,pieplow2024quantum,Debroux2021quantum}, and Pb$V$\cite{Trusheim2019lead,Kuate2021calculating, tchernij2018single,Wang2024transform}. The formalism would also apply to color centers in other hosts or other centers in diamond~\cite{Castelletto2014a,Bockstedte2004ab,Kimoto2014fundamentals,Kuate2018energetics,Gadalla2021enhanced,Kuate2019parallel,Bracher2017selective,Soykal2016silicon,Soykal2017quantum,Weber2010quantum,Gali2011time,Kuate2016strain,Shiang2015ab,Yonggang2024tuning,Huang2024strain,Awschalom2018quantum,Wolfowicz2021quantum,Whiteley2019spin,Kuate2021methods,Day2023laser,Xie2024uncertainty,Chandrasekran2023high,Tandon2024evaluating,Guo2024quantum,Hao2023first}.

This work is organized as follows. We begin by presenting our computational tools in Section \ref{sec:methodology}. Next, our theoretical approach and results for carrier capture are presented in Section \ref{sec:methods} and our proposal for the means of initializing the N$V$ center's charge state and our formalism for the determination of the condition for equilibrium between a collection of point defects in a wide-bandgap semiconductor are provided in Section~\ref{sec:surface}. Finally, we present our conclusions in Section \ref{sec:conc}.

\section{Computational Methods}\label{sec:methodology}
To calculate formation energies for isolated substitutional N defects (N$_\text{C}$) and nitrogen-vacancy centers (N$V$) in diamond, we employed VASP~\cite{Kresse1993ab,Kresse1996efficient,Kuate2021theor,Kresse1999from} with the screened HSE06 hybrid functional for exchange and correlation using the original parameters~\cite{Heyd,Krukau}. Our calculations were terminated when, in the atomic-position relaxations, the forces dropped below a threshold of $10^{-2}$ eV$\cdot$\AA$^{-1}$. We expanded the wavefunctions in a planewave basis with a cutoff energy of 430~eV and employed a 512-atom supercell size ($4\times4\times4$ multiple of the conventional unit cell). The integration scheme used to evaluate the energies was $\Gamma$-point integration. The elements used in our calculations and their associated ground-state structures and chemical-potential values are: N ($\beta$ hexagonal close-packed structure, $-11.39$ eV/atom) and C (diamond structure, $-11.28$ eV/atom). Formation energies were only computed for transitions between charged-defect structures that had been fully relaxed.
 
\section{Theoretical Approach and Results} 
\subsection{Carrier capture}
\label{sec:methods} 
For the tunneling of electrons ($e$) or holes ($h$) between two defects in a manner mediated by the band edges of the crystal, we had derived the expression~\cite{Kuate2023theor}
\begin{align}
\label{eq:ehGamma}
    \bar{\Gamma}_{e,h} &= \sum_{\text{D}~\in~\text{Donors}}N_{\text{D}}\int_{V_\text{D}}\text{d}\mathbf{r}~\rho_\text{D}(\mathbf{r})\sum_{\text{A}~\in~\text{Acceptors}}N_{\text{A}}\int_{V_\text{A}}\text{d}\mathbf{r}^\prime~\rho_\text{A}(\mathbf{r}^\prime)\frac{1}{N_s}\sum_{s}V_{pc}\int\frac{\text{d}\mathbf{k}(t_0)}{(2\pi)^3}\int\frac{\text{d}\mathbf{k}^\prime}{(2\pi)^3}\\&\int_{t_0}^\infty\text{d}t~\frac{1}{t-t_0}\frac{1}{\hbar}\frac{|\nabla_\mathbf{k}\epsilon^\text{C,V}_{\mathbf{k}(t),s}\cdot(\mathbf{r}-\mathbf{r}^\prime)|}{|\mathbf{r}-\mathbf{r}^\prime|^2}\int_{t_0}^t\text{d}t^\prime~\frac{1}{\hbar}\frac{\nabla_\mathbf{k}\epsilon^\text{C,V}_{\mathbf{k}(t^\prime),s}\cdot(\mathbf{r}-\mathbf{r}^\prime)}{|\mathbf{r}-\mathbf{r}^\prime|^2}\nonumber\\&\times\frac{\delta\left(1-\frac{\int_{t_0}^t\text{d}\tilde{t}\frac{1}{\hbar}\nabla_\mathbf{k}\epsilon^\text{C,V}_{\mathbf{k}(\tilde{t}),s}\cdot(\mathbf{r}-\mathbf{r}^\prime)}{|\mathbf{r}-\mathbf{r}^\prime|^2}\right)}{\exp((\epsilon^\text{C,V}_{\mathbf{k}(t_0),s}-E_{\text{F}_\text{D}})/k_BT)+1}\nonumber\times\frac{\delta(\mathbf{k}^\prime-\mathbf{k}(t_0)\mp\frac{1}{\hbar}\int_{t_0}^t\mathbf{F}_{e,h}(\tilde{t}^\prime)\text{d}\tilde{t}^\prime)}{\exp((E_{\text{F}_\text{A}}-\epsilon^\text{C,V}_{\mathbf{k}^\prime,s})/k_BT)+1}\nonumber,
\end{align}
where $\bar{\Gamma}_{e,h}$ is the effective rate for the transfer for electrons or holes, $N_\text{X}$ is the total number of X defects (donors or acceptors) in the entire crystal, $\rho_\text{X}(\mathbf{r})$ is the distribution of the defect X in a Wigner-Seitz cell of volume $V_\text{X} = 1/n_\text{X}$ (where $n_\text{X}$ is the average concentration of the defect X in the crystal), $V_{pc}$ is the volume of the primitive cell, $t_0$ is the time at which the transfer of charge is initiated, $\epsilon^\text{C,V}_{\mathbf{k},s}$ is the energy of the lowest conduction band or highest valence band at wavevector $\mathbf{k}$ with spin $s$, $E_{\text{F}_\text{D}}$ is the donor level, $E_{\text{F}_\text{A}}$ is the acceptor level, and $\mathbf{F}_{e,h}$ is the net force acting on the electrons or the holes. The expression is motivated by the relation governing tunneling from side 2 to side 1 of a contact between a metal and a semiconductor~\cite{kaxiras2019quantum},
\begin{equation}
\label{eq:current21}
    I_{2\rightarrow1} = -\sum_{\mathbf{k}\mathbf{k}^\prime}|\mathcal{T}_{\mathbf{k}\mathbf{k}^\prime}|^2\left[1-f_\mathbf{k}^{(1)}\right]f_{\mathbf{k}^\prime}^{(2)},
\end{equation}
where $f_\mathbf{k}^{(1)}$ and $f_{\mathbf{k}^\prime}^{(2)}$ are the Fermi occupation numbers on each side and $\mathcal{T}_{\mathbf{k}\mathbf{k}^\prime}$ is the tunneling matrix element between the relevant single-particle states. Setting the rate $\bar{\Gamma}_{e,h}$ equal to the current $I_{2\rightarrow1}$ up to units of charge and an overall sign, we can identify
\begin{align}
    |\mathcal{T}_{\mathbf{k}^\prime\mathbf{k}(t_0)}|^2 &= \sum_{\text{D}~\in~\text{Donors}}N_{\text{D}}\int_{V_\text{D}}\text{d}\mathbf{r}~\rho_\text{D}(\mathbf{r})\sum_{\text{A}~\in~\text{Acceptors}}N_{\text{A}}\int_{V_\text{A}}\text{d}\mathbf{r}^\prime~\rho_\text{A}(\mathbf{r}^\prime)\frac{1}{N_s}\sum_{s}\frac{1}{N_{\mathbf{k}(t_0)}}\\&\int_{t_0}^\infty\text{d}t~\frac{1}{t-t_0}\frac{1}{\hbar}\frac{|\nabla_\mathbf{k}\epsilon^\text{C,V}_{\mathbf{k}(t),s}\cdot(\mathbf{r}-\mathbf{r}^\prime)|}{|\mathbf{r}-\mathbf{r}^\prime|^2}\int_{t_0}^t\text{d}t^\prime~\frac{1}{\hbar}\frac{\nabla_\mathbf{k}\epsilon^\text{C,V}_{\mathbf{k}(t^\prime),s}\cdot(\mathbf{r}-\mathbf{r}^\prime)}{|\mathbf{r}-\mathbf{r}^\prime|^2}\nonumber\\&\times\delta\left(1-\frac{\int_{t_0}^t\text{d}\tilde{t}\frac{1}{\hbar}\nabla_\mathbf{k}\epsilon^\text{C,V}_{\mathbf{k}(\tilde{t}),s}\cdot(\mathbf{r}-\mathbf{r}^\prime)}{|\mathbf{r}-\mathbf{r}^\prime|^2}\right)\nonumber\times\delta_{\mathbf{k}^\prime,\mathbf{k}(t_0)\pm\frac{1}{\hbar}\int_{t_0}^t\mathbf{F}_{e,h}(\tilde{t}^\prime)\text{d}\tilde{t}^\prime},
\end{align}
where $N_{\mathbf{k}(t_0)}$ is the total number of ${\mathbf{k}(t_0)}$  states and $\delta_{\mathbf{k}^\prime,\mathbf{k}(t_0)\pm\frac{1}{\hbar}\int_{t_0}^t\mathbf{F}_{e,h}(\tilde{t}^\prime)\text{d}\tilde{t}^\prime}$ is a Kronecker delta that is equal to 1 when $\mathbf{k}^\prime = \mathbf{k}(t_0)\pm\frac{1}{\hbar}\int_{t_0}^t\mathbf{F}_{e,h}(\tilde{t}^\prime)\text{d}\tilde{t}^\prime$ and 0 otherwise, all arising in transforming from integration to summation and from performing the relabelling $\mathbf{k}\rightarrow\mathbf{k}^\prime$ and $\mathbf{k}^\prime\rightarrow\mathbf{k}(t_0)$ in Eq. (\ref{eq:current21}).

Given the proximity to neighboring defects, we have found that charge-state decay experiments for ionized color centers in wide-bandgap semiconductors are better described by tunneling directly between defect states without the intermediary of the crystal band edges~\cite{Kuate2023charge}. In that case we can write the effective rate as
\begin{align}
\label{eq:GammakknoT}
\left<\Gamma\right> &\approx\int_{t_0}^\infty\text{d}t~\frac{1}{t-t_0}\frac{\left|\hbar\mathbf{k}(t)\cdot\Delta\mathbf{r}\right|}{m^*|\Delta\mathbf{r}|^2}\int_{t_0}^t\text{d}t^\prime~\frac{\hbar\mathbf{k}(t^\prime)\cdot\Delta\mathbf{r}}{m^*|\Delta\mathbf{r}|^2}\\&\times\frac{\delta\left(1-\frac{\int_{ t_0}^t\text{d}\tilde{t}\hbar\mathbf{k}(\tilde{t})\cdot\Delta\mathbf{r}}{m^*|\Delta\mathbf{r}|^2}\right)}{\exp\left((E_{\text{F}_\text{D}}-E_{\text{F}_\text{A}})/k_BT\right)+1},\nonumber
\end{align}
where $\hbar\mathbf{k}(t)$ is the definite momentum assigned to a transferring electron at each point in time, $m^*$ is the effective mass associated with the electron, and the Fermi-Dirac distribution factor represents the likelihood that the electron has enough initial kinetic energy to overcome the ionizing potential and recover the neutral system. In this case, there is no distribution factor for hole occupation given that the presence of an ionizing potential implies that the ionized state of the donor is favored and that the electron should not fully rebind to the donor that led to the ionization of the acceptor. The lack of spontaneous charge conversion from N$V^0$ to N$V^-$ in the work of Yuan~\textit{et al.}~\cite{Yuan2020charge} for N$V$ centers that were initialized to N$V^0$ can be explained by the fact that our formalism~\cite{Kuate2023charge} implies that the timescale for spontaneous conversion from N$V^0$ to N$V^-$ will be much faster than the shortest time after which the N$V^-$ population was assessed in the Yuan~\textit{et al.} experiment. Furthermore, the steady-state relative populations of N$V^-$ and N$V^0$ in that experiment and the corresponding values of the Fermi level imply the presence of dopants other than substitutional N to which the charge of the N$V^-$ could be stably transferred. 
Here, we can set the rate $\left<\Gamma\right>$ equal to the current $I_{2\rightarrow1}$ (again up to units of charge and an overall sign) and identify
\begin{align}
     |\mathcal{T}_{\mathbf{k}\mathbf{k}^\prime}|^2 &= \int_{t_0}^\infty\text{d}t\frac{1}{t-t_0}\frac{\left|\hbar\mathbf{k}(t)\cdot\Delta\mathbf{r}\right|}{m^*|\Delta\mathbf{r}|^2}\int_{t_0}^t\text{d}t^\prime\frac{\hbar\mathbf{k}(t^\prime)\cdot\Delta\mathbf{r}}{m^*|\Delta\mathbf{r}|^2}\\&\times\delta\left(1-\frac{\int_{ t_0}^t\text{d}\tilde{t}\hbar\mathbf{k}(\tilde{t})\cdot\Delta\mathbf{r}}{m^*|\Delta\mathbf{r}|^2}\right)\delta_{\mathbf{k},\mathbf{k}(t)}\delta_{\mathbf{k}^\prime,\mathbf{k}(t_0)}.\nonumber
\end{align}
This direct-tunneling formalism implies greater charge-state stability for N$V^-$ in the presence of donors with donor levels far from the N$V^-$ to N$V^0$ charge-transition level so that phosphorus donors would result in greater charge-state stability of the N$V^-$ than nitrogen donors, as verified independently by first-principles calculations~\cite{Zou2024influence}. 

The works of Henry and Lang~\cite{Henry1977nonradiative,Shi2015comparative} and of Marcus~\cite{Marcus1993electron,Shi2015comparative} can help elucidate a mechanism for the tunneling process. In those papers, a large lattice vibration or fluctuations of nuclear coordinates were proposed to account for nonradiative capture or electron transfer. In this work, we argue in a similar manner that a large lattice vibration (a superposition of phonon modes) will cause the wavefunctions of the defects states to exhibit sufficient overlap that charge transfer occurs. Indeed, if the charge transfer is induced by phonon modes then the phonon modes supply the requisite kinetic energy of the model from our earlier work~\cite{Kuate2023charge}. In our prior work~\cite{Kuate2023charge}, we had also noted that the phonons would thermalize the electronic degrees of freedom. We had described the electronic motion using two degrees of freedom, namely a radial degree of freedom $r$ and an angular degree of freedom $\theta$. Therefore, by the equipartition theorem, the thermal energy of the electron would be given by $k_BT$. Projecting onto the radial degree of freedom in order to solve analytically for the electron's velocity, the kinetic energy to be overcome is thus reduced by $E_\text{initial} = k_BT\cos^2(\theta)\cdot\text{sgn}(\cos(\theta))$ so that taking the reciprocal of the rate from Eq. (\ref{eq:GammakknoT}) and averaging over $\theta$ yields the timescale
\begin{align}
\label{eq:tildeGammakkT}
    \left<\tau\right> &\approx \frac{1}{\pi}\int_0^\pi\text{d}\theta\Biggl(\frac{1}{\Delta t}\\&\times\frac{1}{\left(\exp\left(\left(E_{\text{F}_\text{D}}-E_{\text{F}_\text{A}}-E_\text{initial}\right)/k_BT\right)+1\right)}\Biggr)^{-1}.\nonumber
\end{align}
In Eq. (\ref{eq:tildeGammakkT}), the time interval appearing within the integral over $\theta$ is $\Delta t = \int_{r_0}^0\text{d}r\left(\frac{dr}{dt}\right)^{-1}$ where $\frac{dr}{dt} = \pm \sqrt{2/m^*\frac{\left(k_BT\cos^2(\theta)-(E_{\text{F}_\text{D}}-E_{\text{F}_\text{A}})\ln(r/r_0)\right)}{1-\ln(r/r_0)}}$ is the radial velocity obtained from integrating the equation of motion for an electron initially experiencing an electrostatic force having $-\left(E_{\text{F}_\text{D}} - E_{\text{F}_\text{A}}\right)/r$ as the value of its radial component with an initial velocity in the radial direction $\left.\frac{dr}{dt}\right|_{t = t_0} = \sqrt{\frac{2k_BT}{m^*}}\cos(\theta)$ and where $r_0$ is the distance between the donor and the acceptor~\cite{Kuate2023charge}. The positive sign in the expression for $\frac{dr}{dt}$ is employed if the initial velocity of the electron is away from the other dopant while the negative sign is employed if the initial velocity of the electron is toward the other dopant.

Through the works of Henry and Lang~\cite{Henry1977nonradiative} and of Marcus~\cite{Marcus1993electron}, the Fermi-Dirac distribution factor can be interpreted as the likelihood that electrons couple to a state that has been constructed through a superposition of phonon excitations. We note that the construction of a phonon state with energy
\begin{equation}
    E_p^{\text{vib}} = \sum_{\mathbf{k^*},l}\left(n_{\mathbf{k^*},p}^{(l)}+\frac{1}{2}\right)\hbar\omega_\mathbf{k^*}^{(l)}
\end{equation}
corresponds to ionic motion, where $l$ is the ``band'' index~\cite{kaxiras2019quantum}, $\omega_\mathbf{k^*}^{(l)}$ is the frequency of a given phonon indexed by $l$ and the wavevector $\mathbf{k^*}$, and $n_{\mathbf{k^*},p}^{(l)}$ is the number of phonons in the state $p$ with the wavevector $\mathbf{k^*}$ and the index $l$.
This ionic motion then induces electron transfer between the acceptor and donor defects when the energy of the phonon state is sufficient to satisfy
\begin{equation}
    E_p^{\text{vib}} = E_{\text{F}_\text{D}}-E_{\text{F}_\text{A}}-E_\text{initial}.
\end{equation}
Electron-transfer rates corresponding to excitation energies larger than $E_p^{\text{vib}} = E_{\text{F}_\text{D}}-E_{\text{F}_\text{A}}-E_\text{initial}$ are exponentially suppressed and, given the isotropic assumption, the polarization of the phonon state is irrelevant. Together, these observations imply the need to consider only a single phonon state in evaluating the rate of charge transfer between two point defects in the isotropic limit. An advantage of our formulation is that it avoids the need to explicitly construct the required phonon state as a superposition of phonon modes.

We now comment on the condition required for the electronic chemical potential (the Fermi level) to reach equilibrium between two point defects. The definition of chemical equilibrium is that the rate of the forward reaction must be equal to the rate of the reverse reaction. Equivalently, given some initialization of a system, the system has reached equilibrium when it has equal probability of being in the reactant state and transitioning to the product state or in the product state and transitioning to the reactant state, which is to say that we treat the system as ergodic. Explicitly, if 
\begin{align}
\label{eq:tildeGammakkTAD}
    \left<\tau_{\text{A}\rightarrow\text{D}}\right> &\approx \frac{1}{\pi}\int_0^\pi\text{d}\theta\Biggl(\frac{1}{\Delta t}\\&\times\frac{1}{\left(\exp\left(\left(E_{\text{F}_\text{D}}-E_{\text{F}_\text{A}}-E_\text{initial}\right)/k_BT\right)+1\right)}\Biggr)^{-1}\nonumber
\end{align}
is the timescale of charge transfer of an electron from an ionized acceptor to an ionized donor so that the corresponding timescale of charge transfer of an electron from a neutral donor to a neutral acceptor is
\begin{align}
\label{eq:tildeGammakkTDA}
    \left<\tau_{\text{D}\rightarrow\text{A}}\right> &\approx \frac{1}{\pi}\int_0^\pi\text{d}\theta\Biggl(\frac{1}{\Delta t}\\&\times\frac{1}{\left(\exp\left(\left(E_{\text{F}_\text{A}}-E_{\text{F}_\text{D}}+E_\text{initial}\right)/k_BT\right)+1\right)}\Biggr)^{-1},\nonumber
\end{align}
then the maximum expected requisite time for equilibrium to have been reached is 
\begin{equation}
\label{eq:expeq}
    \left<\tau_{\text{equilibrium}}\right> = \left<\tau_{\text{A}\rightarrow\text{D}}\right>+\left<\tau_{\text{D}\rightarrow\text{A}}\right>.
\end{equation}
The result in Eq. (\ref{eq:expeq}) represents the time required to explore all of the relevant phase space, which provides the worst-case timescale in a search to discover equilibrium. In obtaining the timescale for the process $\text{D}\rightarrow\text{A}$, we replace the electron occupation ($f$) from the $\text{A}\rightarrow\text{D}$ process with the hole occupation ($1-f$), where $f$ is the same Fermi-Dirac distribution in both cases. Employing the same Fermi-Dirac distribution in both cases follows from the fact that the binding to the donor can be neglected so that just as the process $\text{A}\rightarrow\text{D}$ did not require a factor ($1-\tilde{f}$), the process $\text{D}\rightarrow\text{A}$ will not require the factor $\tilde{f}$ for some Fermi-Dirac distribution $\tilde{f}$ corresponding to electronic occupation of a donor level.

\begin{figure}[ht!] 
\centering
\includegraphics[width=0.49\textwidth]{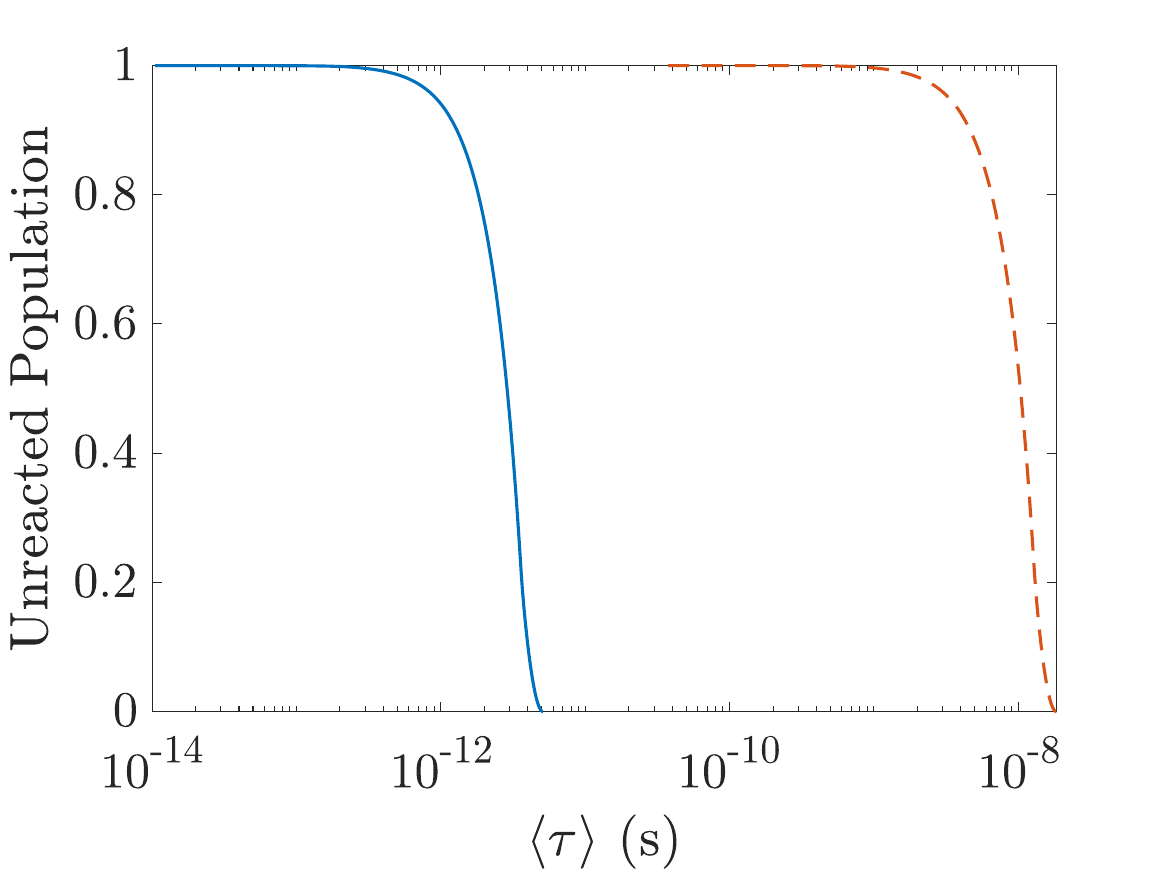}
\caption{Fraction of unreacted population remaining as a function of time, calculated with Eqs. (\ref{eq:tildeGammakkTAD}) and (\ref{eq:tildeGammakkTDA}) and setting $T = 300$~K. We model charge-transfer reactions from an ionized ether-like surface defect to N$_\text{C}^+$ (orange dashed curve) and from N$_\text{C}^0$ to a neutral ether-like surface defect (blue solid curve), where N$_\text{C}$ denotes substitutional N. The ionizing dopant concentration is $n_{\text{N}_\text{C}} = \frac{5\times10^8~\text{cm}^{-2}}{d_{\text{max}}}$, using an estimated maximum implantation depth of $d_\text{max} = 3.5~\text{nm}\cdot\text{keV}^{-1} E_\text{imp}$ where $E_\text{imp}$ is the implantation energy~\cite{Broadway2018spat} ($E_\text{imp} = 3$~keV~\cite{Yuan2020charge}).} 
\label{fig:forwardback}
\end{figure}

The preceding observation can be used to justify the transformation $E_{\text{F}_\text{D}}\rightarrow (E_{\text{F}_\text{D}}+3.4~\text{eV})/2$ from our earlier work~\cite{Kuate2023charge}. The value 3.4~eV is the surface level associated with an ether-like termination of the diamond surface measured relative to the valence band maximum~\cite{Sque2006structure,Broadway2018spat,Zheng2001oxygen}. In Ref.~\cite{Zheng2001oxygen}, the authors apply the LDA method and introduce a GW correction to obtain the value reported herein of 3.4 eV for the surface state. Furthermore, the authors also perform an experiment to determine the location of the surface state and find that it occurs at approximately 3 eV, reported to a precision of one digit, in good agreement with their theoretical calculation. As other authors have demonstrated, however, the determination of the precise chemical makeup of oxygen-terminated diamond is difficult~\cite{Dontschuk2023xray}. Given the agreement of the experimental result in Ref.~\cite{Zheng2001oxygen} with their theoretical value of 3.4 eV for the energy level of the surface state of oxygen-terminated diamond, however, we use the value of 3.4 eV in this work. If we compute the expected timescale $\left<\tau_{\text{A}\rightarrow\text{D}}\right>$ for charge transfer between an ionized acceptor (the ether-like surface level in diamond) and an ionized donor (a substitutional N in diamond)~\cite{Kuate2023charge}, we find that using the nitrogen concentration for sample A in the work of Yuan~\textit{et al.}~\cite{Kuate2023charge,Yuan2020charge} the timescale after which all transitions would be expected to have occurred is approximately 18~ns (see Fig. \ref{fig:forwardback}). In Fig. \ref{fig:forwardback}, the fraction of unreacted population is computed by intersecting a ball of radius $r_0$ with a tetragonal cell of height $d_\text{max}$ that contains on average a single donor for values of $r_0$ ranging from $r_0 = a = 3.549$~\AA~\cite{Kuate2023theor,Kuate2023charge} to $r_0 = \sqrt{\frac{l_{\text{N}_\text{C}}^3}{4d_\text{max}}+d^2_\text{max}}$ where $a$ is the lattice constant of the conventional unit cell of diamond and $l_{\text{N}_\text{C}} = n_{\text{N}_\text{C}}^{-1/3}$, as outlined in Eq. (13) of Ref.~\cite{Kuate2023charge}. For completeness, Ref.~\cite{Sque2006structure}, which applied the LDA method with an energy scaling (though not a GW correction) to fit $I_\text{bulk}-\chi_{\text{bulk}}$ to the experimental bandgap for the clean (001)-oriented surface of diamond, obtained the value $I_\text{bulk} - \chi_{\text{surf}} = (8.21 - 4.66)~\text{eV} = 3.55~\text{eV}$ for the ether-like termination surface state (see Table II of Ref.~\cite{Sque2006structure}).

\begin{figure}[ht!] 
\centering
\includegraphics[width=0.49\textwidth]{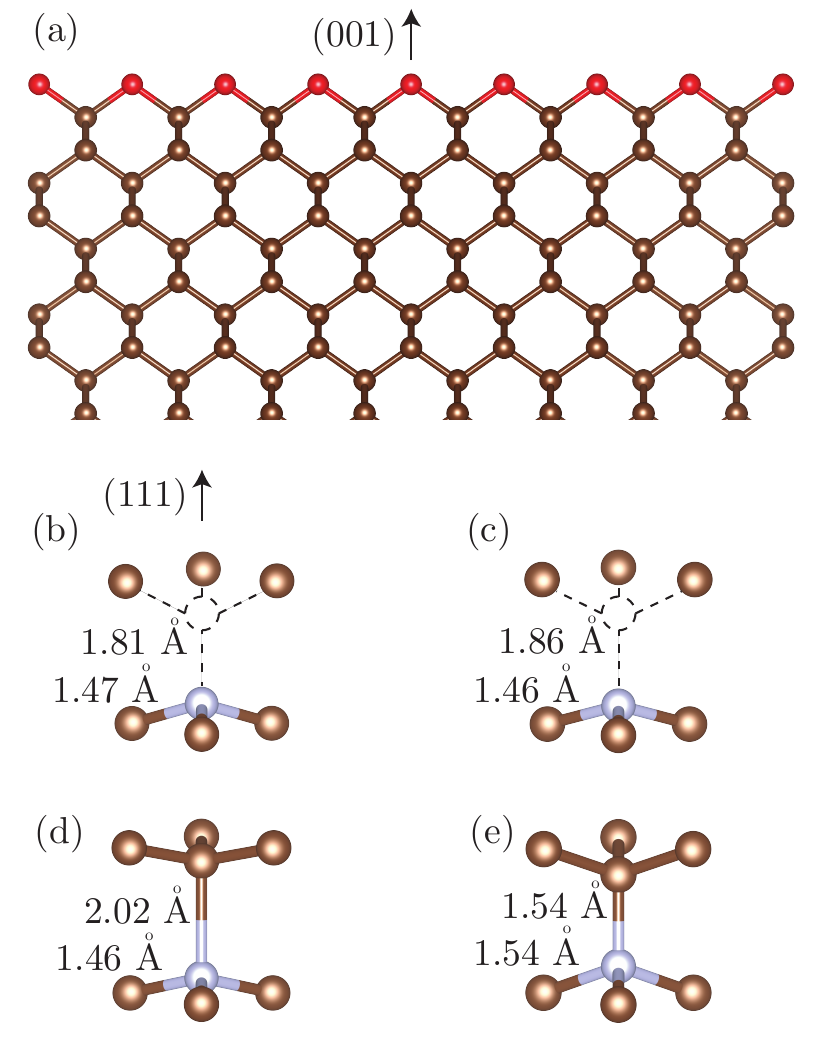}
\caption{Structure of (a) the unrelaxed (001) diamond surface with an ether-like termination, and the relaxed (b) N$V^0$, (c) N$V^-$, (d) N$_\text{C}^0$, and (e) N$_\text{C}^+$ point defects, where N$_\text{C}$ denotes substitutional N. Carbon atoms are shown in brown, nitrogen atoms are shown in purple, and oxygen atoms are shown in red. Carbon vacancies are shown as dashed circles. We obtained the distance between the C vacancy ($V$) and the N atom in the N$V$ defects as described in previous work~\cite{Kuate2021theor}. Defects (b)-(e) have the orientation indicated in (b).} 
\label{fig:structures}
\end{figure}

By contrast, again using the nitrogen concentration for sample A in the work of Yuan~\textit{et al.}~\cite{Kuate2023charge,Yuan2020charge}, we find that the expected timescale $\left<\tau_{\text{D}\rightarrow\text{A}}\right>$ of charge transfer between a neutral donor (a substitutional N in diamond) and a neutral acceptor (the ether-like surface level in diamond)~\cite{Kuate2023charge} for all donor-acceptor pairs in the sample is approximately 5.1~ps (see Fig. \ref{fig:forwardback}). Therefore, for that system, $\left<\tau_{\text{equilibrium}}\right> \approx 18$~ns. The resolution of the photoluminescence (PL) readout was 128 ps averaged over every 100 consecutive points to produce a 12.8~ns averaging window. Thus, over the duration of the capture of a single effective point in the PL trace, no more than on average a single ether-like surface level would have reached equilibrium with a substitutional N level leading to the transformation $E_{\text{F}_\text{D}}\rightarrow (E_{\text{F}_\text{D}}+3.4~\text{eV})/2$~\cite{Kuate2023theor,Kuate2023charge} where $E_{\text{F}_\text{D}}$ is the substitutional N donor level for a perfectly isolated substitutional N defect. By contrast, the minimum timescale for equilibration of all N$V$ centers with substitutional N donors was approximately 2.4~s~\cite{Kuate2023charge}. Therefore, N$V$ centers would experience an environment in which at most pairs of substitutional N and ether-like surface defects had reached equilibrium with an effective donor level given by the transformation $E_{\text{F}_\text{D}}\rightarrow (E_{\text{F}_\text{D}}+3.4~\text{eV})/2$, which produces excellent agreement with experimental results~\cite{Kuate2023charge,Yuan2020charge} (see Fig. \ref{fig:structures} for the structures of the point defects and of the ether-like surface termination).

\subsection{Charge-state initialization and equilibrium condition for point defects}
\label{sec:surface}
In order to employ the charges of N$V$ centers for computation, a reliable procedure for initializing them and a mechanism for transforming charge into a good quantum number must be determined. Leveraging the results from our earlier work~\cite{Kuate2023charge}, we find that if $\tau(p_0)$ is the timescale for charge-state decay of a particular ionized point defect to a relative fraction of the ionized point defect state of $p_0 = 0.5$ and $p$ is actual relative fraction to which the charge state of the point defect has decayed, then $\tau(p) \approx \frac{(1-p)}{p}\tau(p_0)$. Thus, by leaving the point defect under dark conditions for a period of time equal to $\tau(p)$, it is possible to initialize the point defect to any relative fraction $p > 0$ of the ionized point defect state. We next observe that the charge state of a negatively ionized dopant, which would tend to give up its electron, remains a good quantum number if the level associated with the negatively charged dopant is sufficiently far below (much more than $k_BT$) the level of any dopant that would lead to recombination. A similar argument holds for establishing the charge state of a positively ionized dopant as a good quantum number. Therefore, based on the formalism outlined above, modulating the surface termination so that the surface levels are much higher than the donor level of substitutional N would decelerate the charge-state decay of the N$V$ center and allow for halting of the initialization of the N$V$ center's charge state and corresponding halting of computations as well. In a similar vein, adsorbing or desorbing atoms from the surface would accelerate or decelerate the charge-state decay rate and introducing biasing through STM tips brought close to the surface of the crystal sample would have a similar effect. Indeed, recent results~\cite{Rodgers2024diamond} suggest that increased fluorine coverage of hydrogen-terminated diamond shifts the surface level toward the conduction band minimum, which could aid in decelerating the process of charge-state decay once the desired initialization has been achieved, up to the point where the substitutional N donor level becomes farther from the fluorine-induced surface level than from the N$V$ acceptor level. Once the charge state has been initialized, readout of the charge-state could be performed by leveraging the charge-state dependent electroluminescence of the N$V$ center~\cite{Mizuochi2012}.   

In general, the procedure for determining the equilibrium timescale of a collection of point defects proceeds as follows. One would first compute $\left<\tau_{\text{equilibrium}}\right>$ for each pair of defects using Eq. (\ref{eq:expeq}). Next, the pair of defects with the smallest $\left<\tau_{\text{equilibrium}}\right>$ would be treated as a single effective defect and would be assigned an effective charge-transition level given by the average of the charge-transition levels of the defects in the pair. The position of the single effective defect would be chosen such that for that position, which we'll label $r^\prime$ measured along the line from one defect A to the second defect D, $\left<\tau_{\text{A}\rightarrow\text{D}}\right>$ evaluated for $r_0 = r^\prime$ would be equal to $\left<\tau_{\text{D}\rightarrow\text{A}}\right>$ evaluated for $r_0 = r_{\text{A}\rightarrow\text{D}} - r^\prime$, where $r_{\text{A}\rightarrow\text{D}}$ is the distance between the defects A and D. In the subsequent iteration, $\left<\tau_{\text{equilibrium}}\right>$ would be recomputed for each effective defect pair. If the number of physical point defects constituting the pair of effective defects is greater than 2, then the effective charge-transition level for the pair of effective defects with the smallest $\left<\tau_{\text{equilibrium}}\right>$ would be set to the Fermi-level value obtained from Ref.~\cite{Freysoldt2014first}, where the concentration~\cite{Kuate2023theor} of each species is set equal to the number of physical point defects of that species within the pair of effective defects up to an inconsequential constant factor to account for units. Otherwise, the effective charge-transition level for the defect pair would be given by the average of the original charge-transition levels of the defects. In order to obtain the total expected requisite time for equilibrium to have been reached for a collection of $N$ point defects, summation of the expected timescales for equilibrium at each stage in establishing equilibrium within the entire collection of $N$ point defects must be performed. Thus, the expected timescale $\left<T_{\text{equilibrium}}\right>$ for equilibrium of the collection of $N$ point defects is $\left<T_{\text{equilibrium}}\right> = \sum_{i}^\prime\left<\tau^i_{\text{equilibrium}}\right>$, where the terms in the restricted sum iterate through the timescales for the sequence of effective defects of increasing numbers of physical point defects that ultimately culminates in equilibrium within the entire collection of $N$ point defects in the least time. In this manner, vacancies~\cite{Neethirajan2023controlled} and other defects occurring in the dopant formation process could be considered. Thus, given a population of $N$ point defects in thermodynamic equilibrium, the charge state of that population could be initialized by allowing that population to decay in the presence of an additional dopant or an additional effective dopant for the length of time prescribed above, generalizing the notion of donor-acceptor pair quantum technologies~\cite{Bilgin2024donor}.

\section{Conclusions} \label{sec:conc}
In conclusion, we have established the conditions under which the charge-transfer formalisms of Refs.~\cite{Kuate2023charge,Kuate2023theor} are applicable and have shown that the charge state of shallow N$V$ centers in diamond can be efficiently initialized for use as a quantum science platform with charge interfacing enabled by the N$V$ center's charge-state dependent electroluminescence. Additionally, we have provided the condition dictating when point-defect species can be considered to be in equilibrium with one another in order to aid in the determination of the extent of modulation of dopant levels.

\section*{ ACKNOWLEDGMENTS:}
R.K.D. gratefully acknowledges financial support that made this work possible from the Princeton Presidential Postdoctoral Research Fellowship and from the National Academies of Science, Engineering, and Medicine Ford Foundation Postdoctoral Fellowship program. The authors are pleased to acknowledge that the work reported on in this paper was substantially performed using the Princeton Research Computing resources at Princeton University which is consortium of groups led by the Princeton Institute for Computational Science and Engineering (PICSciE) and Office of Information Technology's Research Computing. We additionally acknowledge support by the NSF Science and Technology Center for Integrated Quantum Materials, NSF Grant No. DMR-1231319. Finally, we wish to acknowledge insightful suggestions from Nathalie P. de Leon and we also thank the referees for their many critical and helpful suggestions which have been instrumental in improving the clarity of our paper.
% *****************************************************

\bibliography{refs_NV}
\end{document}